\documentclass[aps,twocolumn,showpacs,preprintnumbers,superscriptaddress,floatfix,nofootinbib]{revtex4-1}

\usepackage[utf8]{inputenc}
\usepackage{slashed}
\usepackage{soul}
\usepackage{indentfirst}
\usepackage{overpic}
\usepackage{color}

\usepackage{amsmath,amssymb,amsfonts}
\usepackage{mathrsfs}
\usepackage{stmaryrd,latexsym,bm,bbm}
\usepackage{epsfig,graphics}
\usepackage{epstopdf}

\usepackage{changes}
\usepackage{threeparttable}

\usepackage[
colorlinks=true,
linkcolor=black,
breaklinks=true,
urlcolor=blue,
citecolor=green]{hyperref}

\usepackage{booktabs}
\usepackage{dcolumn}
\usepackage{makecell}
\usepackage{multirow}

\definecolor{darkblue}{rgb}{0.,0.,0.7}
\definecolor{light-blue}{rgb}{0.8,0.85,1}
\definecolor{green}{rgb}{0,0.6,0}
\definecolor{blueviolet}{rgb}{0.541, 0.169, 0.886}
\definecolor{fuchsia}{rgb}{1.0, 0, 1.0}

\newcommand{\Lag}{\mathcal{L}}

\newcommand{\mev}{\mathrm{MeV}}
\newcommand{\gev}{\mathrm{GeV}}

\begin{document}
	
	\title{Reanalysis of the newly observed $\Omega^*$ state in hadronic molecule model}
	
	\author{Yong-Hui Lin}\email{linyonghui@mail.itp.ac.cn}
	\affiliation{CAS Key Laboratory of Theoretical Physics, Institute
		of Theoretical Physics,  Chinese Academy of Sciences, Beijing 100190, China}
	\affiliation{School of Physical Sciences, University of Chinese Academy
		of Sciences, Beijing 100049, China}

	\author{Fei Wang}\email{wangfei715@mails.ucas.ac.cn}
	\affiliation{School of Physical Sciences, University of Chinese Academy
		of Sciences, Beijing 100049, China}
	
	\author{Bing-Song Zou}\email{zoubs@mail.itp.ac.cn}
	\affiliation{CAS Key Laboratory of Theoretical Physics, Institute
		of Theoretical Physics,  Chinese Academy of Sciences, Beijing 100190, China}
	\affiliation{School of Physical Sciences, University of Chinese Academy
		of Sciences, Beijing 100049, China}
	\affiliation{School of Physics, Central South University, Changsha 410083, China}

	\begin{abstract}
		After the discovery of the new $\Omega^{*}$ state, the ratio of the branching fractions of $\Omega(2012)\to \bar{K}\pi\Xi$ relative to $\bar{K}\Xi$ decay channel was investigated by the Belle Collaboration recently. The measured $11.9\%$ up limit on this ratio is in sharp tension with the $S$-wave $\bar{K}\Xi(1530)$ molecule interpretation for $\Omega(2012)$ which indicates the dominant $\bar{K}\pi\Xi$ three-body decay. In the present work, we try to explore the possibility of the $P$-wave molecule assignments for $\Omega(2012)$ (where $\Omega(2012)$ has positive parity). It is found that the latest experimental measurements are compatible with the $1/2^+$ and $3/2^+$ $\bar{K}\Xi(1530)$ molecular pictures, while the $5/2^+$ $\bar{K}\Xi(1530)$ molecule shows the larger $\bar{K}\pi\Xi$ three-body decay compared with the $\bar{K}\Xi$ decay as the case of $S$-wave molecule. Thus, the newly observed $\Omega(2012)$ can be interpreted as the $1/2^+$ or $3/2^+$ $\bar{K}\Xi(1530)$ molecule state according to current experiment data.
		
	\end{abstract}
	
\maketitle

\section{Introduction}~\label{sec:1}
Last year, the Belle Collaboration reported a new $\Omega^*$ state in the $\bar{K}\Xi$ invariant mass distribution via the $\Upsilon(1S,~2S,~3S)$ decays, with measured mass $M=2012.4 \pm 0.7\ \text{(stat)}\pm 0.6\ \text{(syst)}\ \mev$ and decay width $\Gamma=6.4^{+2.5}_{-2.0}\ \text{(stat)}\pm 1.6\ \text{(syst)}\ \mev$~\cite{Yelton:2018mag}. This discovery aroused bright-eyed interest of theoreticians in understanding the nature of the excited $\Omega$ state. On the one hand, before the discovery of $\Omega(2012)$, only two states, the four-star ground state $\Omega(1672)$ and the three-star excited state $\Omega(2250)$, are listed in the review of the Particle Data Group~(PDG)~\cite{Tanabashi:2018oca} for the $S=-3$ baryon spectrum. In addition, two other two-star states with even higher masses are also mentioned in PDG. The ground state $\Omega(1672)$ is well established in the well-known quark model based on the $SU(3)$-flavor symmetry~\cite{GellMann:1962xb,Neeman:1961jhl}. However, our knowledge on the nature of the $\Omega(2250)$ and other higher $\Omega^*$ states is quite scarce. In particular, the almost $600\ \mathrm{MeV}$ mass difference between the ground state $\Omega(1672)$ and the first observed excited state $\Omega(2250)$ is surprising since the negative-parity orbital excitations of many other baryons are approximately $300\ \mathrm{MeV}$ above their respective ground states. And various models, such as quark model~\cite{Isgur:1978xj,Capstick:1986bm}, Skyrme model~\cite{Oh:2007cr} and lattice gauge theory~\cite{Engel:2013ig}, predicted the masses of the first orbital~($1P$) excitations of $\Omega(1672)$ with $J^P=1/2^-$ or $3/2^-$ are around $2000\ \mev$ which is quite close to the observed value. Stimulated by these facts on the $\Omega$ baryon spectrum, recent theoretical works interpreted the newly observed $\Omega(2012)$ as the $1P$ orbital excitation of the ground state $\Omega$ baryon and investigated its strong decays via the chiral quark model~\cite{Xiao:2018pwe}, QCD sume rules~\cite{Aliev:2018syi,Aliev:2018yjo}, $SU(3)$ flavor symmetry~\cite{Polyakov:2018mow} and ${}^3P_0$ model~\cite{Wang:2018hmi}. The measured width $\sim 6\ \mev$ will be almost saturated by the two-body $\bar{K}\Xi$ channel if the $\Omega(2012)$ is treated as the spin-parity-$3/2^-$ $1P$ excited state. From the point of view of the compact decuplet baryon interpretation for $\Omega(2012)$, the troubling mass gap in the current $S=-3$ baryon spectrum will disappear naturally and the $\Delta(1700)$ resonance with the quantum number of $J^P=3/2^-$ can be assigned as the decuplet partner of $\Omega(2012)$ as pointed out in Ref.~\cite{Polyakov:2018mow}.

On the other hand, the reported mass of the newly observed $\Omega(2012)$ state lies quite close to the threshold of $\bar{K}\Xi(1530)$ channel with the binding energy $\sim 15\ \mev$. The closeness to the thresholds leads naturally to the interpretation of hadronic molecule composed of $\bar{K}\Xi(1530)$ for $\Omega(2012)$. After the first successful attempt of the hadronic molecule picture on uncovering the composite nature of deuteron performed by Weinberg~\cite{Weinberg:1962hj,Weinberg:1965zz}, many exotic states which are discovered during last decades can be described well with the molecule scenarios, such as the $D^*_{s0}(2317)$ as a $DK$ molecule, $X(3872)$ as $D^*\bar{D}$ molecule and newly observed series of $P_c$ pentaquark-like states as $\bar{D}^{(*)}\Sigma_c^{(*)}$ molecules. The systematic discussions on hadronic molecule can be found in the recent reviews~\cite{Guo:2017jvc,Ali:2017jda,Chen:2016qju}. The possibility of $\Omega(2012)$ as the $3/2^-$-$\bar{K}\Xi(1530)$ hadronic molecule is investigated in Refs.~\cite{Polyakov:2018mow,Valderrama:2018bmv,Lin:2018nqd,Huang:2018wth,Pavao:2018xub}. It is remarkable that the three-body $\bar{K}\pi\Xi$ channel contributes sizable width to $\Omega(2012)$ in the $S$-wave hadronic molecule scenario while it is difficult to happen in the compact decuplet baryon assignment. Inspired by this significant difference, the Belle Collaboration tried to distinguish these two interpretations of the $\Omega(2012)$ by searching for its three-body decay to $\bar{K}\pi\Xi$~\cite{Jia:2019eav}. They did not observe the significant $\Omega(2012)$ signals in the $\bar{K}\pi\Xi$ channel and drew their conclusion with the $90\%$ credibility level upper limits on the ratios of the branching fractions of three-body $\bar{K}\pi\Xi$ to $\bar{K}\Xi$ two-body decay, $\mathcal{B}(\Omega(2012)\to\bar{K}\pi\Xi)/\mathcal{B}(\Omega(2012)\to\bar{K}\Xi)<11.9\%$. It is in sharp tension with the prediction of the $S$-wave $\bar{K}\Xi(1530)$ molecule assignment for $\Omega(2012)$. 

However, it should be noted that the quantum number of the $\Omega(2012)$ cannot be determined in experiments at present. Refocusing the processes considered by the Belle Collaboration, the samples of the $\Omega(2012)$ was collected in the final states of the $\Upsilon$ decay processes. The parity of the $\Omega(2012)$ prefers to be positive if the $\Omega(2012)$ is measured via the reaction $\Upsilon\to\bar{\Omega}\Omega^*$. In that way, the $\Upsilon$ decays in $S$ wave which indicates that more events of $\Omega(2012)$ can be observed in experiments. The positive-parity $\Omega(2012)$ should be assigned as the $P$-wave $\bar{K}\Xi(1530)$ bound state in the molecular scenario. Such the existence of the $P$-wave and even higher partial wave bound states in the hidden charm sector was already suggested with the unitary coupled-channel approaches in Ref.~\cite{Shen:2017ayv}. In the present work, we would like to explore the possibility of the $P$-wave molecule assignments for $\Omega(2012)$. Similar to our previous work, we investigate the strong decays of $\Omega(2012)$ with the Effective lagragian approach by treating it as the $P$-wave molecular states with $J^P=1/2^+$, $3/2^+$ and $5/2^+$, respectively.

This work is organized as follows: In Sec.~\ref{sec:2}, we introduce formalism and some details about the theoretical tools used to calculate the decay modes of exotic hadronic molecular states. In Sec.~\ref{sec:3}, the numerical results and discussion are presented. The last section is devoted to the summary of the present work.
\section{Formalism}~\label{sec:2}
\begin{figure*}[htbp]
	\begin{center}
		\includegraphics[width=18cm]{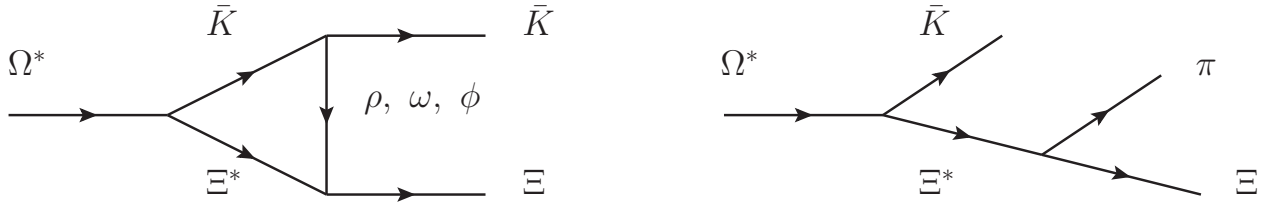}
		\caption{The decay mechanisms of the $\Omega(2012)$ molecules. The left diagram stands for the $\bar{K}\Xi$ two-body decay and the right one is the $\bar{K}\pi\Xi$ three-body decay. $\Omega^*$ and $\Xi^*$ denote the $\Omega(2012)$ and $\Xi(1530)$, respectively.\label{Fig:decay-mechanism}}
	\end{center}
\end{figure*}

In the hadronic molecule picture, the two-body $\bar{K}\Xi$ decay happens via the triangle mechanism where the vector-meson-exchange potential is adopted for the interaction between the $\bar{K}\Xi(1530)$ molecular component and $\bar{K}\Xi$ final state. And the $\bar{K}\pi\Xi$ three-body decay through the decay of the intermediate $\Xi(1530)$ happens in the tree level. The decay diagrams of the $\Omega(2012)$ molecules are shown in the Fig.~\ref{Fig:decay-mechanism}.
The partial decay widths of these diagrams can be calculated with the effective Lagrangian approach. As we did previously, the Lorentz covariant $L$-$S$ scheme proposed in Ref.~\cite{Zou:2002yy} is used to describe the first vertex that $\Omega(2012)$ couples to the $\bar{K}\Xi(1530)$ component in the $P$ wave. The Lagrangians for the different spin parities of the $\Omega(2012)$ are presented in the following,
\begin{align}
\Lag_{\bar{K} \Xi^* \Omega^*(1/2^+)} &= g_{\bar{K} \Xi^* \Omega^*}^{1/2^+} \tilde{g}_{\mu\nu}\left(\partial^\nu\bar{\Xi}^{*\mu}K-\bar{\Xi}^{*\mu}\partial^\nu K\right)\Omega^*,\label{eq:lagrangian12p}\\
\Lag_{\bar{K} \Xi^* \Omega^*(3/2^+)} &= g_{\bar{K} \Xi^* \Omega^*}^{3/2^+} i\epsilon_{\mu\nu\alpha\beta}\tilde{g}^{\alpha\rho}\left(\partial_\rho\bar{\Xi}^{*\nu}K-\bar{\Xi}^{*\nu}\partial_\rho K\right)\notag\\
&\hspace{20mm}\hat{p}^\beta\Omega^{*\mu},\label{eq:lagrangian32p}\\
\Lag_{\bar{K} \Xi^* \Omega^*(5/2^+)} &= g_{\bar{K} \Xi^* \Omega^*}^{5/2^+} \tilde{g}^{\nu\alpha}\left(\partial_\alpha\bar{\Xi}^{*\mu}K-\bar{\Xi}^{*\mu}\partial_\alpha K\right)\Omega^*_{\mu\nu}	
\label{eq:lagrangian52p}
\end{align}
with $\tilde{g}^{\mu\nu}$ defined as $(g^{\mu\nu}-p^\mu p^\nu/p^2)$, where $p$ denotes the momentum of initial $\Omega^*$ state and $\hat{p}=p/m_{\Omega^*}$. The effective couplings $g_{\bar{K} \Xi^* \Omega^*}^{1/2^+}$, $g_{\bar{K} \Xi^* \Omega^*}^{3/2^+}$ and $g_{\bar{K} \Xi^* \Omega^*}^{5/2^+}$ are estimated with the compositeness criterion which states the relation between the derivative of self-energy operator of hadron resonance and its compositeness~\cite{Weinberg:1962hj,Weinberg:1965zz}. In our molecule scenario where the $\Omega(2012)$ is assumed to be the pure $\bar{K}\Xi(1530)$ molecular state, the compositeness of $\Omega(2012)$ equals to one, that is $\chi\equiv1-Z=1$. In general, the compositeness criterion is used to estimate the effective coupling only for the $S$-wave state. It is because only the self-energy loop of the $S$-wave composite state can be evaluated model-independently while for the higher partial wave state, the loop integral is definitely divergent, see Ref.~\cite{Guo:2017jvc} for the detail statements. Consequently, additional scale parameters which are usually the cutoffs in some regulators need to be introduced to cope with the UV divergence in that case. In the present work, we use the compositeness criterion to estimate the $P$-wave couplings between the $\Omega(2012)$ and $\bar{K}\Xi(1530)$ channel by including a Gaussian form factor in the evaluation of the self-energy operator of $\Omega(2012)$. The cutoff dependence of these couplings will be also given when we present our numerical results. The left vertices in the decay diagrams are the same as the $S$-wave case and we take the same convention with our previous calculation~\cite{Lin:2018nqd}.

Finally, two form factors are also included in the loop integrals of the triangle diagrams. The first one is the Gaussian form factor. As discussed in the Ref.~\cite{Lin:2019qiv}, there are two different Gaussian formulas are used commonly in the phenomenological analysis, the four dimensional Euclidean formula~\cite{Faessler:2007gv,Dong:2009yp,Dong:2009tg,Lu:2016nnt,Xiao:2019mst} and the three dimensional nonrelativistic formula~\cite{Nieves:2012tt,HidalgoDuque:2012pq,Guo:2017jvc}. They are defined as
\begin{equation}
f_1(p^2_E /\Lambda_0^2) = {\rm{exp}}(-p^2_E /\Lambda_0^2),
\label{eq:regulator4}
\end{equation}
and
\begin{equation}
f_2(\bm{p}^2 /\Lambda_0^2) = {\rm{exp}}(-\bm{p}^2 /\Lambda_0^2),
\label{eq:regulator3}
\end{equation}
respectively. $p_E$ is the four dimensional Euclidean Jacobi momentum defined as ${m_{\bar{K}}}p_{\Xi^*}/({m_{\bar{K}}+m_{\Xi^{*}}})-{m_{\Xi^*}}p_{\bar{K}}/({m_{\bar{K}}+m_{\Xi^{*}}})$ and $\bm p$ is the spatial part of the momentums of $\bar{K}$ and $\Xi^{*}$ in the rest frame of $\Omega(2012)$ state. Comparing these two equations, we can find that the form factor $f_1$ includes an additional constraint on the energy of molecular components, which demands that the center of mass energy is divided as the mass distribution of compounding particles inside the molecular states as happening usually for the bound states in quantum mechanics. The difference between these two kinds of Gaussian form factors will be presented in the next section.

The second form factor is chosen to be the multipolar formula as shown in Eq.~\eqref{eq:multipolar}. It is introduced to suppress the off-shell contributions of the exchanged mesons in our triangle diagrams.
\begin{equation}
f_3(q^2) = \frac{\Lambda_1^4}{(m^2 - q^2)^2 + \Lambda_1^4},
\label{eq:multipolar}
\end{equation}
where $m$ and $q$ is the mass and momentum of the exchanged particle. The cutoffs $\Lambda_0$ and $\Lambda_1$ are the free parameters in our calculation and we vary both of them in the range of $0.6$-$1.4 \ \mathrm{GeV}$ to scrutinize how the decay behaviors undergo changes as the cutoffs are varied. A specific set of values for $\Lambda_0$ and $\Lambda_1$ is chosen to give the decay patterns of $\Omega(2012)$ molecules by fitting to the measured total widths.

\section{Numerical Results and discussions}~\label{sec:3}
In this section, we will present our numerical results on the strong decays of the $\Omega(2012)$ in the $P$-wave $\bar{K}\Xi^*$ molecule scenarios. Firstly, the short discussion on the estimation of the $P$-wave effective couplings $g_{\Omega^*\bar{K}\Xi^*}$ will be given. After that, we will show the numerical decay patterns of these $P$-wave $\bar{K}\Xi^*$ molecules and also the parameter dependence of our results. The following is the comparison with the latest experimental data and our conclusion. At the end, we try to estimate the partial widths of the three-body $\bar{K}\pi\Xi$ decays which are generated from the rescattering of $\bar{K}\Xi$ channel for the $\Omega(2012)$ states with various quantum numbers.

\subsection{Couplings}
\begin{figure}[htbp]
	\begin{center}
		\includegraphics[width=9cm]{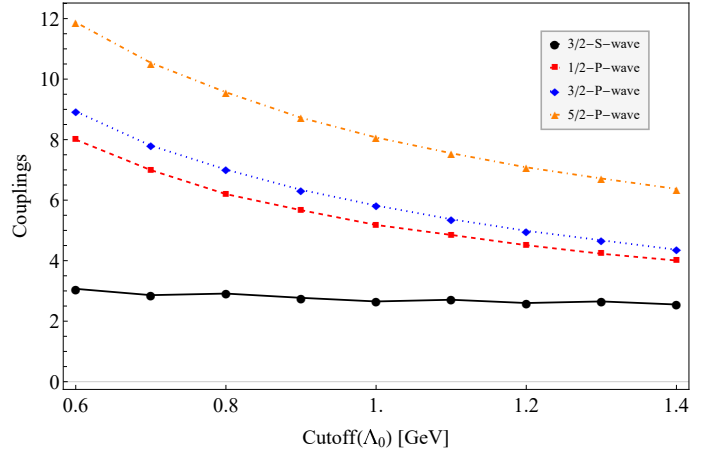}
		\caption{The cutoff dependence of the effective coupling constants $g_{\Omega^*\bar{K}\Xi^*}$ for the different quantum numbers of $\Omega^*$. The black-dotted, red-square, blue-diamond and orange-triangular points denote the cases of the $J^P=3/2^-$ $S$-wave, $1/2^+$ $P$-wave, $3/2^+$ $P$-wave and $5/2^+$ $P$-wave $\Omega^*$ molecules, respectively. \label{Fig:coupling-lambda}}
	\end{center}
\end{figure}

As introduced previously, the effective couplings between the $\Omega(2012)$ and $\bar{K}\Xi^*$ channel are estimated with the compositeness condition. We include the form factor $f_1$~(Eq.~\eqref{eq:regulator4}) in the calculations of the self-energy operators for the $P$-wave $\Omega(2012)$ molecules to get rid of the UV divergence. And for consistently, the same form factor is also included in the estimation of the $S$-wave coupling. The dependence of these effective couplings for various quantum numbers on cutoff $\Lambda_0$ is presented in Fig.~\ref{Fig:coupling-lambda}.
It does not escape attention that the $P$-wave effective couplings $g_{\Omega^*\bar{K}\Xi^*}^{1/2^+}$, $g_{\Omega^*\bar{K}\Xi^*}^{3/2^+}$ and $g_{\Omega^*\bar{K}\Xi^*}^{5/2^+}$ are more sensitive to the cutoff $\Lambda_0$ than the $S$-wave coupling $g_{\Omega^*\bar{K}\Xi^*}^{3/2^-}$ as expected. And all the couplings decrease when the cutoff gets larger. The $5/2^+$ $\Omega(2012)$ has the largest coupling with the $\bar{K}\Xi^*$ channel, $1/2^+$ and $3/2^+$ are next, and the $3/2^-$ is smallest. In particular, $g_{\Omega^*\bar{K}\Xi^*}^{1/2^+}$ is quite similar with  $g_{\Omega^*\bar{K}\Xi^*}^{3/2^+}$. In spite of some model-dependence existence, the uncertainty of the determination of $P$-wave effective couplings from the compositeness condition is still under control. As shown in Fig.~\ref{Fig:coupling-lambda}, the largest magnitude of the coupling decrease is less than the half of its value in the whole range of $\Lambda_0$. It is the authors' opinion that the $P$-wave effective couplings obtained with the compositeness condition are available for the estimation of the decay widths of the molecular states, especially to estimate the relative ratios of branch fractions among various decay channels.

\subsection{Partial decay widths of the $\Omega(2012)$ molecules}
With the couplings $g_{\Omega^*\bar{K}\Xi^*}$ obtained, the partial widths of the $\Omega^*$ molecules can be calculated straightforwardly. The results with the cutoff $\Lambda_0=1.0\ \gev$ and $\Lambda_1=0.8\ \gev$ which are fitted to the measured width of $\Omega(2012)$ are displayed in Table~\ref{table:widthrt} for the form factor set $(f_1, f_3)$ and Table~\ref{table:widthnr} for the $(f_2, f_3)$.
\begin{table}[htpb]
	\centering
	\caption{\label{table:widthrt}Partial widths of $\Omega(2012)$ as the $\bar K \Xi(1530)$ molecules with various quantum numbers. And the cutoffs are fixed as $\Lambda_0=1.0\ \mathrm{GeV}$, $\Lambda_1=0.8\ \mathrm{GeV}$. The form factor set ($f_1$, $f_3$) is used in the calculation. All of the decay widths are in the unit of $\mathrm{MeV}$.}
	\begin{tabular}{l|*{4}{c}}
		\Xhline{1pt}
		\multirow{4}*{Mode} & \multicolumn{4}{c}{Widths ($\mathrm{MeV}$) with ($f_1$, $f_3$)} \\
		\Xcline{2-5}{0.4pt}
		& \multicolumn{4}{c}{$\Omega(2012)$($\bar K\Xi(1530)$)} \\
		\Xcline{2-5}{0.4pt}
		& \multicolumn{1}{c}{$S$-wave}& \multicolumn{3}{c}{$P$-wave}  \\
		\Xcline{2-5}{0.4pt}
		& ${\frac32}^-$& ${\frac12}^+$ & ${\frac32}^+$ & ${\frac52}^+$ \\
		\Xhline{0.8pt}
		$\bar K\Xi$ 	  	 &0.05 &6.6 &4.4 &0.005 \\
		$\bar K\pi\Xi$ 	  	 &1.9 &0.2 &0.2 &0.3 \\
		\Xhline{0.8pt}
		Total 				 &1.95 &6.8 &4.6 &0.305\\
		Ratio	  	 &38 &0.03 &0.045 &60 \\
		\Xhline{1pt}
	\end{tabular}
\end{table}
\begin{table}[htpb]
	\centering
	\caption{\label{table:widthnr}Partial widths of $\Omega(2012)$ as the $\bar K \Xi(1530)$ molecules with various quantum numbers. And the cutoffs are fixed as $\Lambda_0=1.0\ \mathrm{GeV}$, $\Lambda_1=0.8\ \mathrm{GeV}$. The form factor set ($f_2$, $f_3$) is used in the calculation. All of the decay widths are in the unit of $\mathrm{MeV}$.}
	\begin{tabular}{l|*{4}{c}}
		\Xhline{1pt}
		\multirow{4}*{Mode} & \multicolumn{4}{c}{Widths ($\mathrm{MeV}$) with ($f_2$, $f_3$)} \\
		\Xcline{2-5}{0.4pt}
		& \multicolumn{4}{c}{$\Omega(2012)$($\bar K\Xi(1530)$)} \\
		\Xcline{2-5}{0.4pt}
		& \multicolumn{1}{c}{$S$-wave}& \multicolumn{3}{c}{$P$-wave}  \\
		\Xcline{2-5}{0.4pt}
		& ${\frac32}^-$& ${\frac12}^+$ & ${\frac32}^+$ & ${\frac52}^+$ \\
		\Xhline{0.8pt}
		$\bar K\Xi$ 	  	 &0.4 &22.7 &91.2 &0.1 \\
		$\bar K\pi\Xi$ 	  	 &1.9 &0.2 &0.2 &0.3 \\
		\Xhline{0.8pt}
		Total 				 &2.3 &22.9 &91.4 &0.4\\
		Ratio	  	 &4.75 &0.009 &0.002 &3 \\
		\Xhline{1pt}
	\end{tabular}
\end{table}
It is intriguing that the $3/2^-$ $S$-wave and $5/2^+$ $P$-wave $\Omega(2012)$ molecules have quite similar decay patterns except the highly suppressed widths of $5/2^+$ state and the decay pattern of the $1/2^+$ state is almost same with that of the $3/2^+$ state. And the remarkable difference between these two set of states is that the dominant decay channel of $1/2^+$ and $3/2^+$ $P$-wave states is the two-body $\bar{K}\Xi$ channel which is compatible with the experimental observations, while it is the three-body $\bar{K}\pi\Xi$ channel for the $3/2^-$ $S$-wave and $5/2^+$ $P$-wave states which is in sharp tension with the experimental data. It also can be noticed that the form factor $f_2$ gives much larger decay width for the $\bar{K}\Xi$ channel than $f_1$. It is similar with the case discussed in Ref.~\cite{Lin:2019qiv}, the exchanged vector mesons~(with the energy about $0.2\ \gev$) are off the mass shell when the component particles $\bar{K}$ and $\Xi(1530)$ are confined on nearly their mass shells by the form factor $f_1$ and its contribution will be suppressed by the form factor $f_3$.

\begin{figure*}[htbp]
	\begin{center}
		\includegraphics[width=18cm]{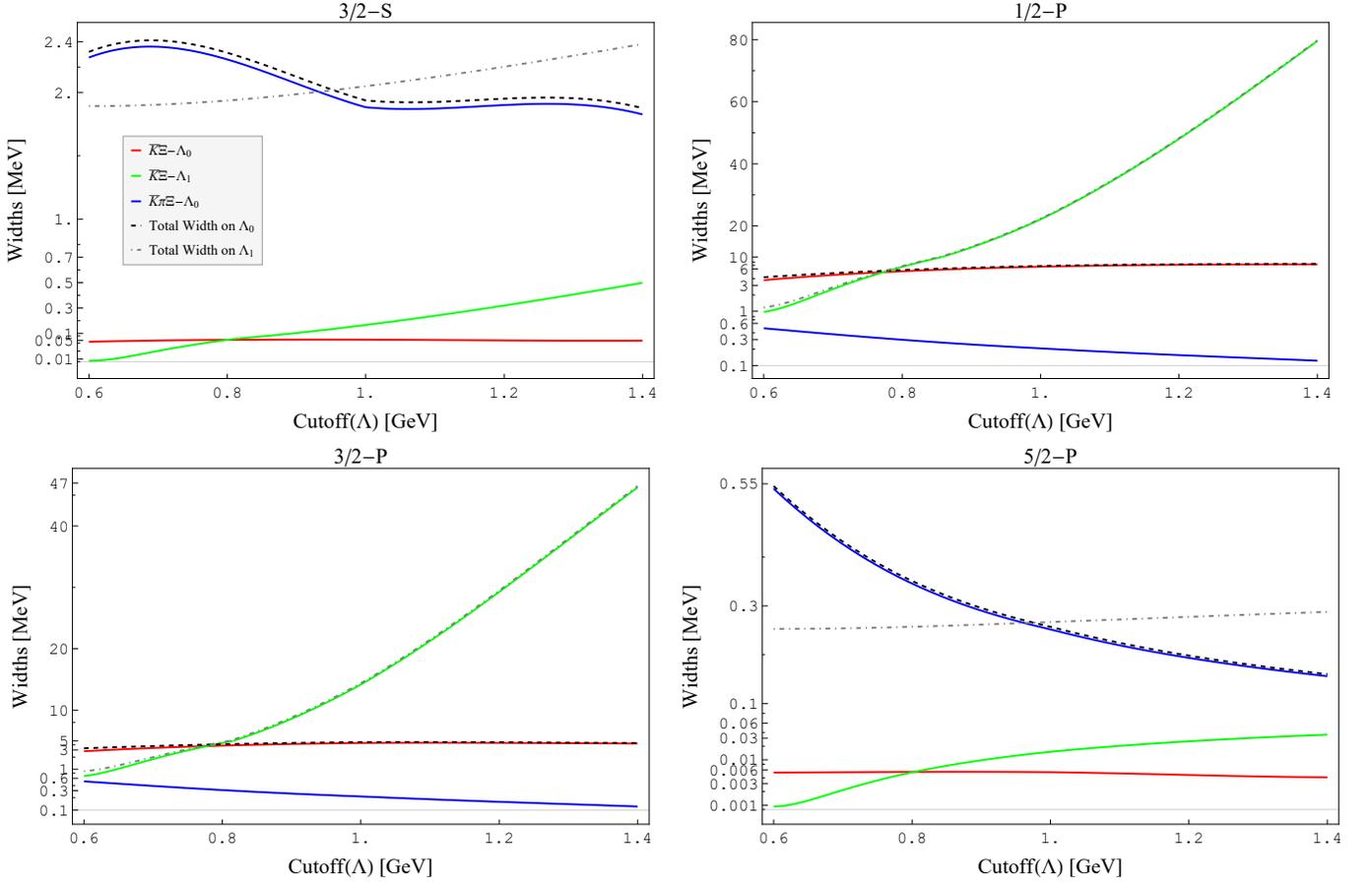}
		\caption{Decay widths of the $\Omega^*$ molecules with various quantum numbers varying with the cutoff $\Lambda_0$ and $\Lambda_1$ obtained with the form factor set of $(f_1, f_3)$. The left upper, left lower, right upper and right lower plots are the results of the $\Omega^*$ molecule with $J^P=3/2^-$, $3/2^+$, $1/2^+$ and $5/2^+$, respectively. The red-solid, green-solid, blue-solid, black-solid and black-dashed lines denote the $\Lambda_0$ dependence of $\bar{K}\Xi$ channel, the $\Lambda_1$ dependence of $\bar{K}\Xi$ channel, the $\Lambda_0$ dependence of $\bar{K}\pi\Xi$ channel, the $\Lambda_0$ dependence of total width and the $\Lambda_1$ dependence of total width, respectively. Note that the $\Lambda_0$ is fixed at $1.0\ \gev$ when $\Lambda_1$ is varied and the $\Lambda_1$ is fixed at $0.8\ \gev$ when $\Lambda_0$ is varied. \label{Fig:widths-cutoff}}
	\end{center}
\end{figure*}

\begin{figure*}[htbp]
	\begin{center}
		\includegraphics[width=18cm]{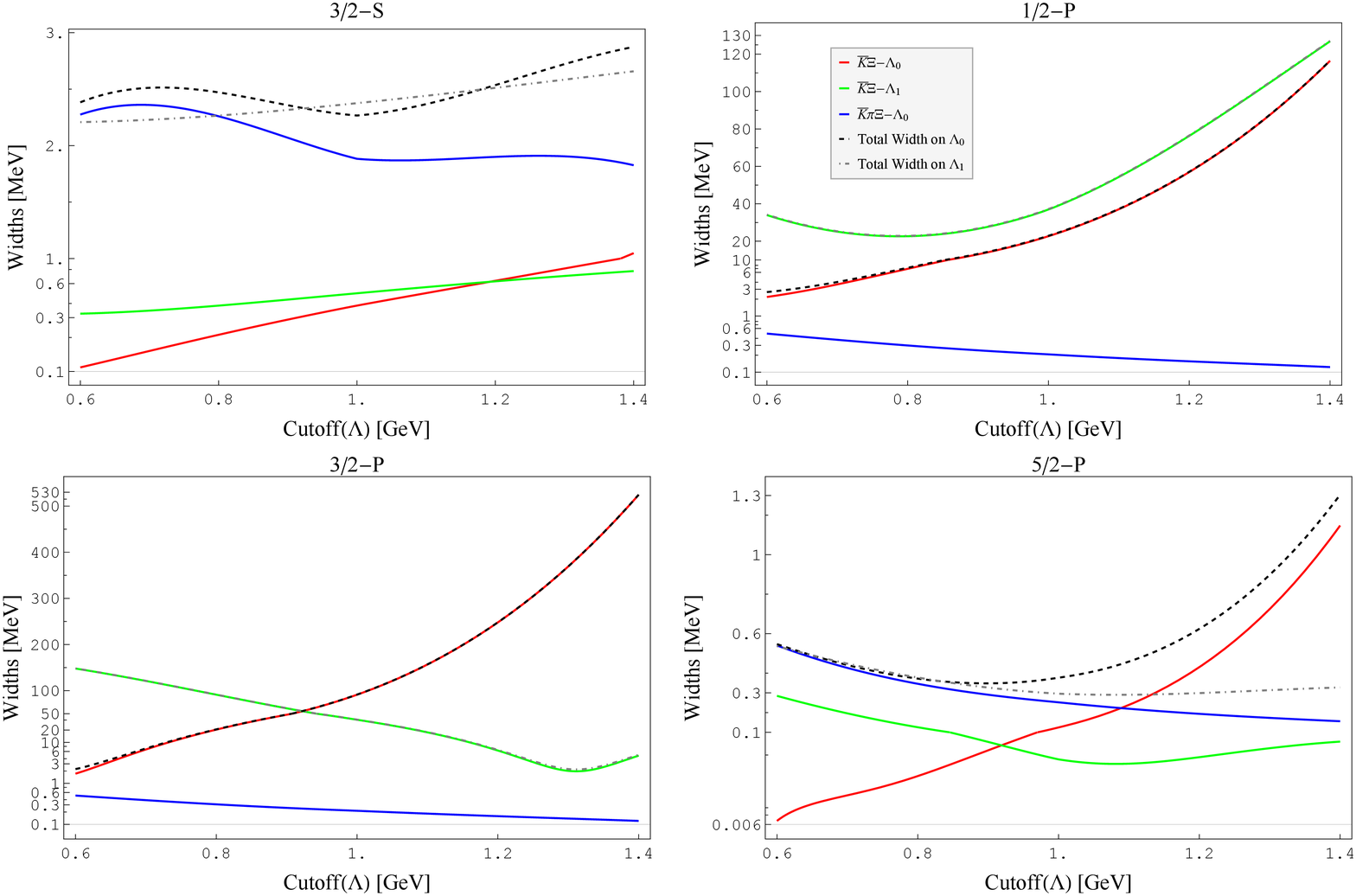}
		\caption{Decay widths of the $\Omega^*$ molecules with various quantum numbers varying with the cutoff $\Lambda_0$ and $\Lambda_1$ obtained with the form factor set of $(f_2, f_3)$. The left upper, left lower, right upper and right lower plots are the results of the $\Omega^*$ molecule with $J^P=3/2^-$, $3/2^+$, $1/2^+$ and $5/2^+$, respectively. The red-solid, green-solid, blue-solid, black-solid and black-dashed lines denote the $\Lambda_0$ dependence of $\bar{K}\Xi$ channel, the $\Lambda_1$ dependence of $\bar{K}\Xi$ channel, the $\Lambda_0$ dependence of $\bar{K}\pi\Xi$ channel, the $\Lambda_0$ dependence of total width and the $\Lambda_1$ dependence of total width, respectively. Note that the $\Lambda_0$ is fixed at $1.0\ \gev$ when $\Lambda_1$ is varied and the $\Lambda_1$ is fixed at $0.8\ \gev$ when $\Lambda_0$ is varied. \label{Fig:widths-cutoff-nr}}
	\end{center}
\end{figure*}

The cutoff dependence of the partial decay widths of $\Omega(2012)$ states with various quantum numbers are given in the Fig.~\ref{Fig:widths-cutoff} for the form factor set $(f_1, f_3)$ and Fig.~\ref{Fig:widths-cutoff-nr} for the $(f_2, f_3)$ case. Also the widths of $P$-wave states are more sensitive to both cutoff $\Lambda_0$ and $\Lambda_1$ than that of the $S$-wave state. The partial width of the three-body $\bar{K}\pi\Xi$ channel only depends slightly on the cutoff $\Lambda_0$ due to the cutoff dependent coupling constant $g_{\Omega^*\bar{K}\Xi}^{3/2^-}$.
The partial widths of the two-body $\bar{K}\Xi$ channel obtained with the form factor set $(f_1, f_3)$ depend a lot on the $\Lambda_1$ while keep almost steady as the $\Lambda_0$ is varied. In the case where the form factor set $(f_2, f_3)$ is used, however, the two-body $\bar{K}\Xi$ decay widths depend heavily on the $\Lambda_0$ and its dependence on $\Lambda_1$ is relatively modest but not such slight as the $\Lambda_0$ dependence in the case of $(f_1, f_3)$.

\begin{figure*}[htbp]
	\begin{center}
		\includegraphics[width=18cm]{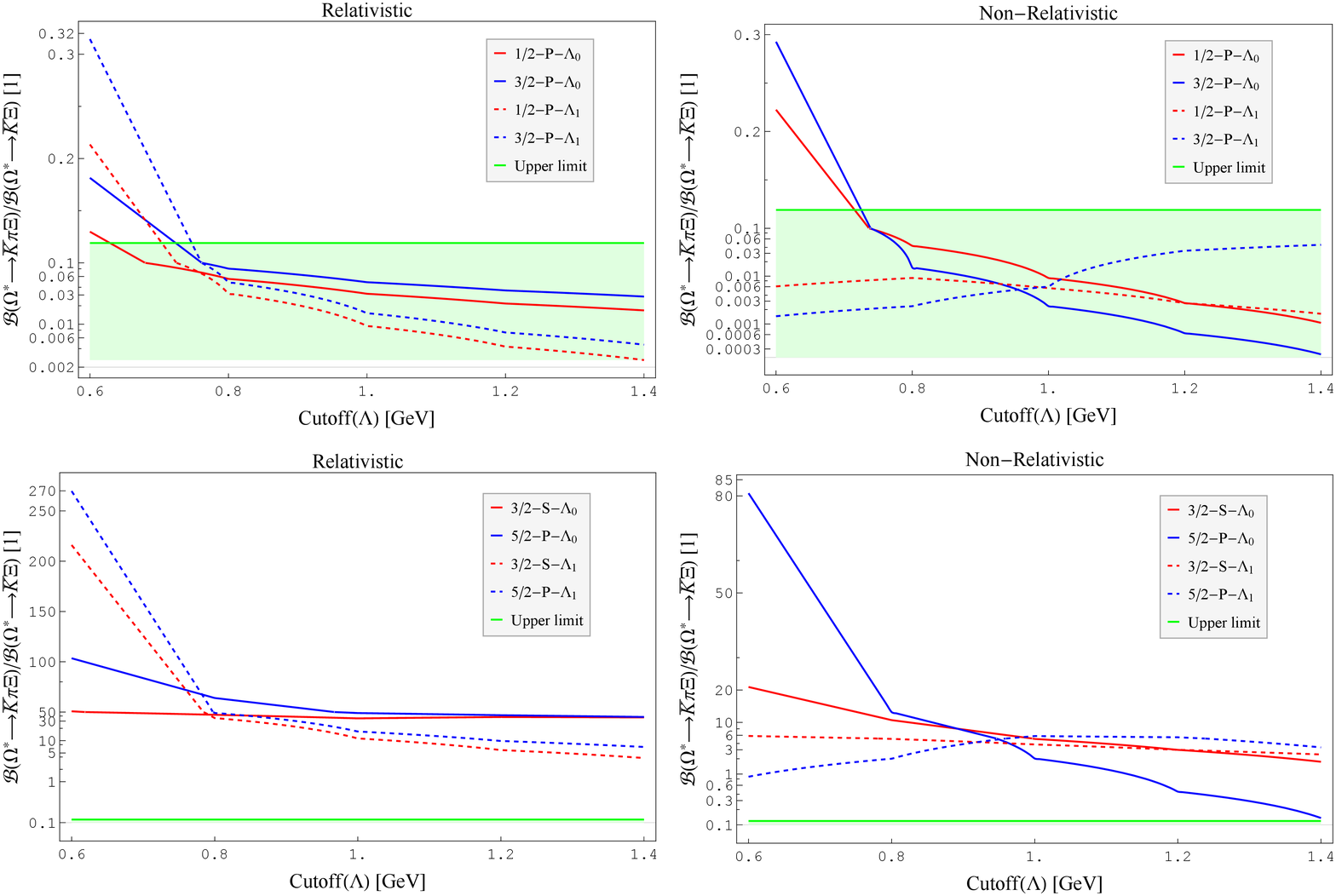}
		\caption{The ratio of the branching fractions of three-body $\bar{K}\pi\Xi$ channel relative to the two-body $\bar{K}\Xi$ channel varying with the cutoff $\Lambda_0$ and $\Lambda_1$. The upper panels are the results of the $\Omega^*$ molecule with $J^P=1/2^+$ and $3/2^+$, where the left one is obtained with the form factor set of $(f_1, f_3)$ and the right is obtained with the form factor set of $(f_2, f_3)$. The lower panels are for the $J^P=3/2^-$ and $5/2^+$ cases. And the green band in the plot denotes the region allowed by the latest Belle measurements. Note that the $\Lambda_0$ is fixed at $1.0\ \gev$ when $\Lambda_1$ is varied and the $\Lambda_1$ is fixed at $0.8\ \gev$ when $\Lambda_0$ is varied.\label{Fig:ratio}}
	\end{center}
\end{figure*}

Besides the decay widths, the dependence of the relative ratios between the partial decay widths of $\bar{K}\pi\Xi$ and $\bar{K}\Xi$ channels are also considered and the results are displayed in Fig.~\ref{Fig:ratio}. Although the decay widths of $\Omega(2012)$ molecules are cutoff dependent, the characteristic behaviors on the ratios of the branching fractions of $\bar{K}\pi\Xi$ channel relative to the $\bar{K}\Xi$ channel of the $\bar{K}\Xi^*$ molecules with different spin parities are deserving of special attention. As shown in Fig.~\ref{Fig:ratio}, these ratios of all the $\Omega(2012)$ molecules does not change significantly in the whole range of cutoffs relative to the experimental upper limit. The green bands in the these plots denote the allowed region of the ratios by experimental data. The spin-parity $1/2^+$ and $3/2^+$ $\bar{K}\Xi(1530)$ molecular assignments for the $\Omega(2012)$ are consistent with the experiments when the cut off is larger than $0.7\ \gev$. While the $3/2^-$ and $5/2^+$ $\bar{K}\Xi(1530)$ molecular assumptions strongly disagree with the experiments in the range of cutoffs we considered. Then it can pave another way to understand the nature of $\Omega(2012)$ state besides the $1P$ orbital excitation of the ground $\Omega$ state. Our results suggest the $\Omega(2012)$ might be the $P$-wave $\bar{K}\Xi(1530)$ molecule state with $J^P=1/2^+$ or $3/2^+$. Since the decay behaviors of the $1/2^+$ and $3/2^+$ $\bar{K}\Xi(1530)$ molecules are quite similar with each other, it is difficult to distinguish these two possible quantum numbers in the current hadronic molecular framework. The experimental determination of the quantum number for the $\Omega(2012)$ state is critical to uncover the mystery on its inner structure in future.

\subsection{The $\bar{K}\pi\Xi$ decays via the rescattering of $\bar{K}\Xi$}
\begin{figure}[htbp]
	\begin{center}
		\includegraphics[width=9cm]{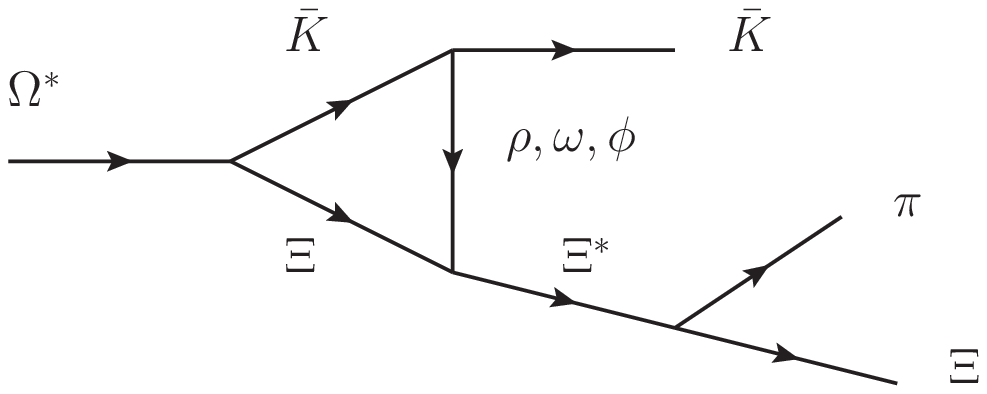}
		\caption{The three-body $\bar{K}\pi\Xi$ decay diagrams of the $\Omega(2012)$ through the rescattering of $\bar{K}\Xi$. $\Omega^*$ and $\Xi^*$ denote the $\Omega(2012)$ and $\Xi(1530)$, respectively.\label{Fig:rescattering}}
	\end{center}
\end{figure}

We also investigate the three-body $\bar{K}\pi\Xi$ decays of $\Omega(2012)$ by considering the $\bar{K}\pi\Xi$ are generated from the rescattering of $\bar{K}\Xi$. The $\Omega(2012)$ states are suspected to decay into $\bar{K}\Xi$ firstly. And next the $\bar{K}\Xi$ rescatters to the $\bar{K}\Xi(1530)$ channel which can be described by the vector meson dominance model. Finally, the three-body $\bar{K}\pi\Xi$ final state is generated through the decay of the intermediate $\Xi(1530)$. The decay mechanism is shown in Fig.~\ref{Fig:rescattering}. Here the quantum numbers of $\Omega(2012)$ are considered to be $1/2^-$, $1/2^+$, $3/2^-$, $3/2^+$, $5/2^-$ and $5/2^+$. We expect that these various quantum numbers of $\Omega^(2012)$ can be distinguished by such three-body $\bar{K}\pi\Xi$ decays. Similarly, the partial widths of these diagrams in Fig.~\ref{Fig:rescattering} are calculated with the effective Lagrangian approach. The interaction between $\Omega(2012)$ and $\bar{K}\Xi$ is extracted from the experimental width of $\Omega(2012)$ by assuming that it is completely saturated by the $\bar{K}\Xi$ channel with the following effective lagrangians,
\begin{align}
\mathcal{L}_{\bar{K}\Xi\Omega^*(1/2^-)}&=g_{\bar{K}\Xi\Omega^*}^{1/2^-}\bar{\Xi}K\Omega^*,\\
\mathcal{L}_{\bar{K}\Xi\Omega^*(1/2^+)}&=g_{\bar{K}\Xi\Omega^*}^{1/2^+}\bar{\Xi}\gamma^5 \tilde{\gamma}^{\mu} r_{\mu} K\Omega^*,\\
\mathcal{L}_{\bar{K}\Xi\Omega^*(3/2^+)}&=g_{\bar{K}\Xi\Omega^*}^{3/2^+}\bar{\Xi}r_{\mu} K\Omega^{*\mu},\\
\mathcal{L}_{\bar{K}\Xi\Omega^*(3/2^-)}&=g_{\bar{K}\Xi\Omega^*}^{3/2^-}\bar{\Xi}\gamma^5\left(\tilde{\gamma}^{\alpha}g^{\beta\mu}+\tilde{\gamma}^{\beta}g^{\alpha\mu}\right)\mathcal{D}_{\alpha\beta}K\Omega^{*}_\mu,\\
\mathcal{L}_{\bar{K}\Xi\Omega^*(5/2^-)}&=g_{\bar{K}\Xi\Omega^*}^{5/2^-}\bar{\Xi}\mathcal{D}^{\mu\nu}K\Omega^{*}_{\mu\nu},\\
\mathcal{L}_{\bar{K}\Xi\Omega^*(5/2^+)}&=g_{\bar{K}\Xi\Omega^*}^{5/2^+}\bar{\Xi}\gamma^5\big(\tilde{\gamma}^{\lambda}g^{\alpha\mu}g^{\beta\nu}+\tilde{\gamma}^{\alpha}g^{\lambda\mu}g^{\beta\nu}\notag\\
&\hspace{10mm}+\tilde{\gamma}^{\beta}g^{\alpha\mu}g^{\lambda\nu}\big)\mathcal{F}_{\alpha\beta\lambda}K\Omega^{*}_{\mu\nu},
\end{align}
where
\begin{align}
\tilde{g}^{\mu\nu}&=\left(g^{\mu\nu}-\frac{p^\mu p^\nu}{m_{\Omega^*}}\right),\notag\\
r^{\mu}&=\tilde{g}^{\mu\nu}\left(k_1-k_2\right)_\nu, \qquad
\tilde{\gamma}^{\mu}=\tilde{g}^{\mu\nu}\gamma_\nu,\notag\\
\mathcal{D}^{\mu\nu}&=r^\mu r^\nu-\frac13 r^\rho r_\rho\tilde{g}^{\mu\nu},\notag\\
\mathcal{F}^{\mu\nu\lambda}&=r^\mu r^\nu r^\lambda-\frac15 r^\rho r_\rho\left(\tilde{g}^{\mu\nu}r^\lambda+\tilde{g}^{\nu\lambda}r^\mu+\tilde{g}^{\lambda\mu}r^\nu\right)\notag,
\end{align}
with $p$, $k_1$, $k_2$ the momentums of $\Omega(2012)$, $\bar{K}$ and $\Xi$, respectively. A single form factor $f_2$ is included to cope with the UV divergence and $\Lambda_0=1.0\ \gev$ is chosen to give a qualitative estimations of the $\bar{K}\pi\Xi$ decays. The extracted coupling constants $g_{\bar{K}\Xi\Omega^*}$ are listed in Table~\ref{table:couplings}.
\begin{table}[htpb]
	\centering
	\caption{\label{table:couplings}The effective coupling constants $g_{\bar{K}\Xi\Omega^*}$ for the $\Omega(2012)$ states with various quantum numbers.}
	\begin{tabular}{l|*{6}{c}}
		\Xhline{1pt}
		\thead{Quantum\\ numbers} & $1/2^-$ & \thead{$1/2^+$ \\ ($\mathrm{GeV}^{-1}$)}& \thead{$3/2^+$ \\ ($\mathrm{GeV}^{-1}$)}& \thead{$3/2^-$ \\ ($\mathrm{GeV}^{-2}$)}& \thead{$5/2^-$ \\ ($\mathrm{GeV}^{-2}$)}& \thead{$5/2^+$ \\ ($\mathrm{GeV}^{-3}$)} \\
		\Xhline{0.4pt}
		$g_{\bar{K}\Xi\Omega^*}$ &0.39 &0.49 &0.85 &0.53 &1.68 &0.71\\
		\Xhline{1pt}
	\end{tabular}
\end{table}
The corresponding partial widths are displayed in Table.~\ref{table:three-body}. The partial widths of $\bar{K}\pi\Xi$ channel for all the $\Omega(2012)$ states are quite small owing to the triangle loop mechanism. The $3/2^-$ $\Omega(2012)$ has the largest decay width to the $\bar{K}\pi\Xi$ channel of around $5.0\times10^{-2}\ \mev$. The $1/2^+$ and $3/2^+$ $\Omega(2012)$ states have similar decay widths of the magnitude of order $10^{-3}\ \mev$ and the three-body partial widths of the left $\Omega(2012)$ states are varied from $10^{-5}$ to $10^{-4}\ \mev$. It is similar with the cases of the $\Omega(2012)$ molecules where the $1/2^+$ and $3/2^+$ $\bar{K}\Xi(1530)$ molecules have the similar three-body decays and the $3/2^-$ $\bar{K}\Xi(1530)$ molecule has much larger three-body decays relatively. Note that these small $\bar{K}\pi\Xi$ partial widths obtained in this subsection are irrelevant with the inner structure of $\Omega(2012)$. They can serve as the lower limits of the partial decay widths of the $\bar{K}\pi\Xi$ channel for the $\Omega(2012)$ state although they cannot help us to discriminate various quantum numbers of $\Omega(2012)$.
\begin{table*}[htpb]
	\centering
	\caption{\label{table:three-body}The partial widths of the three-body $\bar{K}\pi\Xi$ decays for the $\Omega(2012)$ states with various quantum numbers. These values are obtained with $\Lambda_0=1\ \gev$. All of the decay widths are in the unit of $\mathrm{MeV}$.}
	\begin{tabular}{l|*{6}{c}}
		\Xhline{1pt}
		\thead{Quantum\\ numbers} & $1/2^-$ & $1/2^+$& $3/2^+$& $3/2^-$& $5/2^-$& $5/2^+$ \\
		\Xhline{0.4pt}
		$\Gamma_{\bar{K}\pi\Xi}$ &$3.6\times10^{-5}$ &$5.1\times10^{-3}$ &$2.8\times10^{-3}$ &$5.0\times10^{-2}$ &$4.7\times10^{-5}$ &$4.0\times10^{-4}$\\
		\Xhline{1pt}
	\end{tabular}
\end{table*}
\section{Summary}~\label{sec:4}
The latest observation on the ratio of the branching fractions of $\Omega(2012)\to \bar{K}\pi\Xi$ relative to the $\bar{K}\Xi$ channel reported by the Belle Collaboration strongly disfavors the $S$-wave $\bar{K}\Xi(1530)$ molecule interpretation for the $\Omega(2012)$. It seems to indicate that the $\Omega(2012)$ can only be considered as the $1P$ orbital excitation of the ground $\Omega$ baryon with $J^P=3/2^-$. In fact, there is no definite conclusion on the quantum number of the newly observed $\Omega(2012)$ state so far. In the present work, we explore the possibility of the $\Omega(2012)$ being a $P$-wave $\bar{K}\Xi(1530)$ molecular state. Analogous to our previous work, we investigate the strong decays of the $P$-wave $\Omega(2012)$ molecules with the quantum numbers of $1/2^+$, $3/2^+$ and $5/2^+$ by using the effective lagrangian approach. It is found that the decay behaviors of the $1/2^+$ and $3/2^+$ $\Omega(2012)$ molecules are compatible with the current experimental data. Then it suggests that $\Omega(2012)$ might be the $P$-wave $\bar{K}\Xi(1530)$ molecules with $J^P=1/2^+$ or $3/2^+$ besides being the $J^P=3/2^-$ orbital excitation of the ground $\Omega$ baryon. The determination of the quantum number of $\Omega(2012)$ would be a landmark experimental feat on understanding its nature.

\bigskip
\noindent
\begin{center}
	{\bf ACKNOWLEDGEMENTS}\\
	
\end{center}
This project is supported by NSFC under Grant
No. 11621131001 (CRC110 cofunded by DFG and NSFC) and Grant No. 11835015.

\bibliography{Omega-refs}

\begin{thebibliography}{33}%
\makeatletter
\providecommand \@ifxundefined [1]{%
 \@ifx{#1\undefined}
}%
\providecommand \@ifnum [1]{%
 \ifnum #1\expandafter \@firstoftwo
 \else \expandafter \@secondoftwo
 \fi
}%
\providecommand \@ifx [1]{%
 \ifx #1\expandafter \@firstoftwo
 \else \expandafter \@secondoftwo
 \fi
}%
\providecommand \natexlab [1]{#1}%
\providecommand \enquote  [1]{``#1''}%
\providecommand \bibnamefont  [1]{#1}%
\providecommand \bibfnamefont [1]{#1}%
\providecommand \citenamefont [1]{#1}%
\providecommand \href@noop [0]{\@secondoftwo}%
\providecommand \href [0]{\begingroup \@sanitize@url \@href}%
\providecommand \@href[1]{\@@startlink{#1}\@@href}%
\providecommand \@@href[1]{\endgroup#1\@@endlink}%
\providecommand \@sanitize@url [0]{\catcode `\\12\catcode `\$12\catcode
  `\&12\catcode `\#12\catcode `\^12\catcode `\_12\catcode `\%12\relax}%
\providecommand \@@startlink[1]{}%
\providecommand \@@endlink[0]{}%
\providecommand \url  [0]{\begingroup\@sanitize@url \@url }%
\providecommand \@url [1]{\endgroup\@href {#1}{\urlprefix }}%
\providecommand \urlprefix  [0]{URL }%
\providecommand \Eprint [0]{\href }%
\providecommand \doibase [0]{http://dx.doi.org/}%
\providecommand \selectlanguage [0]{\@gobble}%
\providecommand \bibinfo  [0]{\@secondoftwo}%
\providecommand \bibfield  [0]{\@secondoftwo}%
\providecommand \translation [1]{[#1]}%
\providecommand \BibitemOpen [0]{}%
\providecommand \bibitemStop [0]{}%
\providecommand \bibitemNoStop [0]{.\EOS\space}%
\providecommand \EOS [0]{\spacefactor3000\relax}%
\providecommand \BibitemShut  [1]{\csname bibitem#1\endcsname}%
\let\auto@bib@innerbib\@empty
\bibitem [{\citenamefont {Yelton}\ \emph {et~al.}(2018)\citenamefont {Yelton}
  \emph {et~al.}}]{Yelton:2018mag}%
  \BibitemOpen
  \bibfield  {author} {\bibinfo {author} {\bibfnamefont {J.}~\bibnamefont
  {Yelton}} \emph {et~al.} (\bibinfo {collaboration} {Belle}),\ }\href
  {\doibase 10.1103/PhysRevLett.121.052003} {\bibfield  {journal} {\bibinfo
  {journal} {Phys. Rev. Lett.}\ }\textbf {\bibinfo {volume} {121}},\ \bibinfo
  {pages} {052003} (\bibinfo {year} {2018})},\ \Eprint
  {http://arxiv.org/abs/1805.09384} {arXiv:1805.09384 [hep-ex]} \BibitemShut
  {NoStop}%
\bibitem [{\citenamefont {Tanabashi}\ \emph {et~al.}(2018)\citenamefont
  {Tanabashi} \emph {et~al.}}]{Tanabashi:2018oca}%
  \BibitemOpen
  \bibfield  {author} {\bibinfo {author} {\bibfnamefont {M.}~\bibnamefont
  {Tanabashi}} \emph {et~al.} (\bibinfo {collaboration} {Particle Data
  Group}),\ }\href {\doibase 10.1103/PhysRevD.98.030001} {\bibfield  {journal}
  {\bibinfo  {journal} {Phys. Rev.}\ }\textbf {\bibinfo {volume} {D98}},\
  \bibinfo {pages} {030001} (\bibinfo {year} {2018})}\BibitemShut {NoStop}%
\bibitem [{\citenamefont {Gell-Mann}(1962)}]{GellMann:1962xb}%
  \BibitemOpen
  \bibfield  {author} {\bibinfo {author} {\bibfnamefont {M.}~\bibnamefont
  {Gell-Mann}},\ }\href {\doibase 10.1103/PhysRev.125.1067} {\bibfield
  {journal} {\bibinfo  {journal} {Phys. Rev.}\ }\textbf {\bibinfo {volume}
  {125}},\ \bibinfo {pages} {1067} (\bibinfo {year} {1962})}\BibitemShut
  {NoStop}%
\bibitem [{\citenamefont {Ne'eman}(1961)}]{Neeman:1961jhl}%
  \BibitemOpen
  \bibfield  {author} {\bibinfo {author} {\bibfnamefont {Y.}~\bibnamefont
  {Ne'eman}},\ }\href {\doibase 10.1016/0029-5582(61)90134-1} {\bibfield
  {journal} {\bibinfo  {journal} {Nucl. Phys.}\ }\textbf {\bibinfo {volume}
  {26}},\ \bibinfo {pages} {222} (\bibinfo {year} {1961})},\ \bibinfo {note}
  {[,34(1961)]}\BibitemShut {NoStop}%
\bibitem [{\citenamefont {Isgur}\ and\ \citenamefont
  {Karl}(1978)}]{Isgur:1978xj}%
  \BibitemOpen
  \bibfield  {author} {\bibinfo {author} {\bibfnamefont {N.}~\bibnamefont
  {Isgur}}\ and\ \bibinfo {author} {\bibfnamefont {G.}~\bibnamefont {Karl}},\
  }\href {\doibase 10.1103/PhysRevD.18.4187} {\bibfield  {journal} {\bibinfo
  {journal} {Phys. Rev.}\ }\textbf {\bibinfo {volume} {D18}},\ \bibinfo {pages}
  {4187} (\bibinfo {year} {1978})}\BibitemShut {NoStop}%
\bibitem [{\citenamefont {Capstick}\ and\ \citenamefont
  {Isgur}(1986)}]{Capstick:1986bm}%
  \BibitemOpen
  \bibfield  {author} {\bibinfo {author} {\bibfnamefont {S.}~\bibnamefont
  {Capstick}}\ and\ \bibinfo {author} {\bibfnamefont {N.}~\bibnamefont
  {Isgur}},\ }\bibfield  {booktitle} {\emph {\bibinfo {booktitle}
  {{Proceedings, International Conference on Hadron Spectroscopy: College Park,
  Maryland, April 20-22, 1985}}},\ }\href {\doibase 10.1103/PhysRevD.34.2809,
  10.1063/1.35361} {\bibfield  {journal} {\bibinfo  {journal} {Phys. Rev.}\
  }\textbf {\bibinfo {volume} {D34}},\ \bibinfo {pages} {2809} (\bibinfo {year}
  {1986})},\ \bibinfo {note} {[AIP Conf. Proc.132,267(1985)]}\BibitemShut
  {NoStop}%
\bibitem [{\citenamefont {Oh}(2007)}]{Oh:2007cr}%
  \BibitemOpen
  \bibfield  {author} {\bibinfo {author} {\bibfnamefont {Y.}~\bibnamefont
  {Oh}},\ }\href {\doibase 10.1103/PhysRevD.75.074002} {\bibfield  {journal}
  {\bibinfo  {journal} {Phys. Rev.}\ }\textbf {\bibinfo {volume} {D75}},\
  \bibinfo {pages} {074002} (\bibinfo {year} {2007})},\ \Eprint
  {http://arxiv.org/abs/hep-ph/0702126} {arXiv:hep-ph/0702126 [HEP-PH]}
  \BibitemShut {NoStop}%
\bibitem [{\citenamefont {Engel}\ \emph {et~al.}(2013)\citenamefont {Engel},
  \citenamefont {Lang}, \citenamefont {Mohler},\ and\ \citenamefont
  {Schäfer}}]{Engel:2013ig}%
  \BibitemOpen
  \bibfield  {author} {\bibinfo {author} {\bibfnamefont {G.~P.}\ \bibnamefont
  {Engel}}, \bibinfo {author} {\bibfnamefont {C.~B.}\ \bibnamefont {Lang}},
  \bibinfo {author} {\bibfnamefont {D.}~\bibnamefont {Mohler}}, \ and\ \bibinfo
  {author} {\bibfnamefont {A.}~\bibnamefont {Schäfer}} (\bibinfo
  {collaboration} {BGR}),\ }\href {\doibase 10.1103/PhysRevD.87.074504}
  {\bibfield  {journal} {\bibinfo  {journal} {Phys. Rev.}\ }\textbf {\bibinfo
  {volume} {D87}},\ \bibinfo {pages} {074504} (\bibinfo {year} {2013})},\
  \Eprint {http://arxiv.org/abs/1301.4318} {arXiv:1301.4318 [hep-lat]}
  \BibitemShut {NoStop}%
\bibitem [{\citenamefont {Xiao}\ and\ \citenamefont
  {Zhong}(2018)}]{Xiao:2018pwe}%
  \BibitemOpen
  \bibfield  {author} {\bibinfo {author} {\bibfnamefont {L.-Y.}\ \bibnamefont
  {Xiao}}\ and\ \bibinfo {author} {\bibfnamefont {X.-H.}\ \bibnamefont
  {Zhong}},\ }\href {\doibase 10.1103/PhysRevD.98.034004} {\bibfield  {journal}
  {\bibinfo  {journal} {Phys. Rev.}\ }\textbf {\bibinfo {volume} {D98}},\
  \bibinfo {pages} {034004} (\bibinfo {year} {2018})},\ \Eprint
  {http://arxiv.org/abs/1805.11285} {arXiv:1805.11285 [hep-ph]} \BibitemShut
  {NoStop}%
\bibitem [{\citenamefont {Aliev}\ \emph
  {et~al.}(2018{\natexlab{a}})\citenamefont {Aliev}, \citenamefont {Azizi},
  \citenamefont {Sarac},\ and\ \citenamefont {Sundu}}]{Aliev:2018syi}%
  \BibitemOpen
  \bibfield  {author} {\bibinfo {author} {\bibfnamefont {T.~M.}\ \bibnamefont
  {Aliev}}, \bibinfo {author} {\bibfnamefont {K.}~\bibnamefont {Azizi}},
  \bibinfo {author} {\bibfnamefont {Y.}~\bibnamefont {Sarac}}, \ and\ \bibinfo
  {author} {\bibfnamefont {H.}~\bibnamefont {Sundu}},\ }\href {\doibase
  10.1103/PhysRevD.98.014031} {\bibfield  {journal} {\bibinfo  {journal} {Phys.
  Rev.}\ }\textbf {\bibinfo {volume} {D98}},\ \bibinfo {pages} {014031}
  (\bibinfo {year} {2018}{\natexlab{a}})},\ \Eprint
  {http://arxiv.org/abs/1806.01626} {arXiv:1806.01626 [hep-ph]} \BibitemShut
  {NoStop}%
\bibitem [{\citenamefont {Aliev}\ \emph
  {et~al.}(2018{\natexlab{b}})\citenamefont {Aliev}, \citenamefont {Azizi},
  \citenamefont {Sarac},\ and\ \citenamefont {Sundu}}]{Aliev:2018yjo}%
  \BibitemOpen
  \bibfield  {author} {\bibinfo {author} {\bibfnamefont {T.~M.}\ \bibnamefont
  {Aliev}}, \bibinfo {author} {\bibfnamefont {K.}~\bibnamefont {Azizi}},
  \bibinfo {author} {\bibfnamefont {Y.}~\bibnamefont {Sarac}}, \ and\ \bibinfo
  {author} {\bibfnamefont {H.}~\bibnamefont {Sundu}},\ }\href {\doibase
  10.1140/epjc/s10052-018-6375-y} {\bibfield  {journal} {\bibinfo  {journal}
  {Eur. Phys. J.}\ }\textbf {\bibinfo {volume} {C78}},\ \bibinfo {pages} {894}
  (\bibinfo {year} {2018}{\natexlab{b}})},\ \Eprint
  {http://arxiv.org/abs/1807.02145} {arXiv:1807.02145 [hep-ph]} \BibitemShut
  {NoStop}%
\bibitem [{\citenamefont {Polyakov}\ \emph {et~al.}(2019)\citenamefont
  {Polyakov}, \citenamefont {Son}, \citenamefont {Sun},\ and\ \citenamefont
  {Tandogan}}]{Polyakov:2018mow}%
  \BibitemOpen
  \bibfield  {author} {\bibinfo {author} {\bibfnamefont {M.~V.}\ \bibnamefont
  {Polyakov}}, \bibinfo {author} {\bibfnamefont {H.-D.}\ \bibnamefont {Son}},
  \bibinfo {author} {\bibfnamefont {B.-D.}\ \bibnamefont {Sun}}, \ and\
  \bibinfo {author} {\bibfnamefont {A.}~\bibnamefont {Tandogan}},\ }\href
  {\doibase 10.1016/j.physletb.2019.03.054} {\bibfield  {journal} {\bibinfo
  {journal} {Phys. Lett.}\ }\textbf {\bibinfo {volume} {B792}},\ \bibinfo
  {pages} {315} (\bibinfo {year} {2019})},\ \Eprint
  {http://arxiv.org/abs/1806.04427} {arXiv:1806.04427 [hep-ph]} \BibitemShut
  {NoStop}%
\bibitem [{\citenamefont {Wang}\ \emph {et~al.}(2018)\citenamefont {Wang},
  \citenamefont {Gui}, \citenamefont {Lü}, \citenamefont {Xiao},\ and\
  \citenamefont {Zhong}}]{Wang:2018hmi}%
  \BibitemOpen
  \bibfield  {author} {\bibinfo {author} {\bibfnamefont {Z.-Y.}\ \bibnamefont
  {Wang}}, \bibinfo {author} {\bibfnamefont {L.-C.}\ \bibnamefont {Gui}},
  \bibinfo {author} {\bibfnamefont {Q.-F.}\ \bibnamefont {Lü}}, \bibinfo
  {author} {\bibfnamefont {L.-Y.}\ \bibnamefont {Xiao}}, \ and\ \bibinfo
  {author} {\bibfnamefont {X.-H.}\ \bibnamefont {Zhong}},\ }\href {\doibase
  10.1103/PhysRevD.98.114023} {\bibfield  {journal} {\bibinfo  {journal} {Phys.
  Rev.}\ }\textbf {\bibinfo {volume} {D98}},\ \bibinfo {pages} {114023}
  (\bibinfo {year} {2018})},\ \Eprint {http://arxiv.org/abs/1810.08318}
  {arXiv:1810.08318 [hep-ph]} \BibitemShut {NoStop}%
\bibitem [{\citenamefont {Weinberg}(1963)}]{Weinberg:1962hj}%
  \BibitemOpen
  \bibfield  {author} {\bibinfo {author} {\bibfnamefont {S.}~\bibnamefont
  {Weinberg}},\ }\href {\doibase 10.1103/PhysRev.130.776} {\bibfield  {journal}
  {\bibinfo  {journal} {Phys. Rev.}\ }\textbf {\bibinfo {volume} {130}},\
  \bibinfo {pages} {776} (\bibinfo {year} {1963})}\BibitemShut {NoStop}%
\bibitem [{\citenamefont {Weinberg}(1965)}]{Weinberg:1965zz}%
  \BibitemOpen
  \bibfield  {author} {\bibinfo {author} {\bibfnamefont {S.}~\bibnamefont
  {Weinberg}},\ }\href {\doibase 10.1103/PhysRev.137.B672} {\bibfield
  {journal} {\bibinfo  {journal} {Phys. Rev.}\ }\textbf {\bibinfo {volume}
  {137}},\ \bibinfo {pages} {B672} (\bibinfo {year} {1965})}\BibitemShut
  {NoStop}%
\bibitem [{\citenamefont {Guo}\ \emph {et~al.}(2018)\citenamefont {Guo},
  \citenamefont {Hanhart}, \citenamefont {Meißner}, \citenamefont {Wang},
  \citenamefont {Zhao},\ and\ \citenamefont {Zou}}]{Guo:2017jvc}%
  \BibitemOpen
  \bibfield  {author} {\bibinfo {author} {\bibfnamefont {F.-K.}\ \bibnamefont
  {Guo}}, \bibinfo {author} {\bibfnamefont {C.}~\bibnamefont {Hanhart}},
  \bibinfo {author} {\bibfnamefont {U.-G.}\ \bibnamefont {Meißner}}, \bibinfo
  {author} {\bibfnamefont {Q.}~\bibnamefont {Wang}}, \bibinfo {author}
  {\bibfnamefont {Q.}~\bibnamefont {Zhao}}, \ and\ \bibinfo {author}
  {\bibfnamefont {B.-S.}\ \bibnamefont {Zou}},\ }\href {\doibase
  10.1103/RevModPhys.90.015004} {\bibfield  {journal} {\bibinfo  {journal}
  {Rev. Mod. Phys.}\ }\textbf {\bibinfo {volume} {90}},\ \bibinfo {pages}
  {015004} (\bibinfo {year} {2018})},\ \Eprint
  {http://arxiv.org/abs/1705.00141} {arXiv:1705.00141 [hep-ph]} \BibitemShut
  {NoStop}%
\bibitem [{\citenamefont {Ali}\ \emph {et~al.}(2017)\citenamefont {Ali},
  \citenamefont {Lange},\ and\ \citenamefont {Stone}}]{Ali:2017jda}%
  \BibitemOpen
  \bibfield  {author} {\bibinfo {author} {\bibfnamefont {A.}~\bibnamefont
  {Ali}}, \bibinfo {author} {\bibfnamefont {J.~S.}\ \bibnamefont {Lange}}, \
  and\ \bibinfo {author} {\bibfnamefont {S.}~\bibnamefont {Stone}},\ }\href
  {\doibase 10.1016/j.ppnp.2017.08.003} {\bibfield  {journal} {\bibinfo
  {journal} {Prog. Part. Nucl. Phys.}\ }\textbf {\bibinfo {volume} {97}},\
  \bibinfo {pages} {123} (\bibinfo {year} {2017})},\ \Eprint
  {http://arxiv.org/abs/1706.00610} {arXiv:1706.00610 [hep-ph]} \BibitemShut
  {NoStop}%
\bibitem [{\citenamefont {Chen}\ \emph {et~al.}(2016)\citenamefont {Chen},
  \citenamefont {Chen}, \citenamefont {Liu},\ and\ \citenamefont
  {Zhu}}]{Chen:2016qju}%
  \BibitemOpen
  \bibfield  {author} {\bibinfo {author} {\bibfnamefont {H.-X.}\ \bibnamefont
  {Chen}}, \bibinfo {author} {\bibfnamefont {W.}~\bibnamefont {Chen}}, \bibinfo
  {author} {\bibfnamefont {X.}~\bibnamefont {Liu}}, \ and\ \bibinfo {author}
  {\bibfnamefont {S.-L.}\ \bibnamefont {Zhu}},\ }\href {\doibase
  10.1016/j.physrep.2016.05.004} {\bibfield  {journal} {\bibinfo  {journal}
  {Phys. Rept.}\ }\textbf {\bibinfo {volume} {639}},\ \bibinfo {pages} {1}
  (\bibinfo {year} {2016})},\ \Eprint {http://arxiv.org/abs/1601.02092}
  {arXiv:1601.02092 [hep-ph]} \BibitemShut {NoStop}%
\bibitem [{\citenamefont {Valderrama}(2018)}]{Valderrama:2018bmv}%
  \BibitemOpen
  \bibfield  {author} {\bibinfo {author} {\bibfnamefont {M.~P.}\ \bibnamefont
  {Valderrama}},\ }\href {\doibase 10.1103/PhysRevD.98.054009} {\bibfield
  {journal} {\bibinfo  {journal} {Phys. Rev.}\ }\textbf {\bibinfo {volume}
  {D98}},\ \bibinfo {pages} {054009} (\bibinfo {year} {2018})},\ \Eprint
  {http://arxiv.org/abs/1807.00718} {arXiv:1807.00718 [hep-ph]} \BibitemShut
  {NoStop}%
\bibitem [{\citenamefont {Lin}\ and\ \citenamefont {Zou}(2018)}]{Lin:2018nqd}%
  \BibitemOpen
  \bibfield  {author} {\bibinfo {author} {\bibfnamefont {Y.-H.}\ \bibnamefont
  {Lin}}\ and\ \bibinfo {author} {\bibfnamefont {B.-S.}\ \bibnamefont {Zou}},\
  }\href {\doibase 10.1103/PhysRevD.98.056013} {\bibfield  {journal} {\bibinfo
  {journal} {Phys. Rev.}\ }\textbf {\bibinfo {volume} {D98}},\ \bibinfo {pages}
  {056013} (\bibinfo {year} {2018})},\ \Eprint
  {http://arxiv.org/abs/1807.00997} {arXiv:1807.00997 [hep-ph]} \BibitemShut
  {NoStop}%
\bibitem [{\citenamefont {Huang}\ \emph {et~al.}(2018)\citenamefont {Huang},
  \citenamefont {Liu}, \citenamefont {Lu}, \citenamefont {Xie},\ and\
  \citenamefont {Geng}}]{Huang:2018wth}%
  \BibitemOpen
  \bibfield  {author} {\bibinfo {author} {\bibfnamefont {Y.}~\bibnamefont
  {Huang}}, \bibinfo {author} {\bibfnamefont {M.-Z.}\ \bibnamefont {Liu}},
  \bibinfo {author} {\bibfnamefont {J.-X.}\ \bibnamefont {Lu}}, \bibinfo
  {author} {\bibfnamefont {J.-J.}\ \bibnamefont {Xie}}, \ and\ \bibinfo
  {author} {\bibfnamefont {L.-S.}\ \bibnamefont {Geng}},\ }\href {\doibase
  10.1103/PhysRevD.98.076012} {\bibfield  {journal} {\bibinfo  {journal} {Phys.
  Rev.}\ }\textbf {\bibinfo {volume} {D98}},\ \bibinfo {pages} {076012}
  (\bibinfo {year} {2018})},\ \Eprint {http://arxiv.org/abs/1807.06485}
  {arXiv:1807.06485 [hep-ph]} \BibitemShut {NoStop}%
\bibitem [{\citenamefont {Pavao}\ and\ \citenamefont
  {Oset}(2018)}]{Pavao:2018xub}%
  \BibitemOpen
  \bibfield  {author} {\bibinfo {author} {\bibfnamefont {R.}~\bibnamefont
  {Pavao}}\ and\ \bibinfo {author} {\bibfnamefont {E.}~\bibnamefont {Oset}},\
  }\href {\doibase 10.1140/epjc/s10052-018-6329-4} {\bibfield  {journal}
  {\bibinfo  {journal} {Eur. Phys. J.}\ }\textbf {\bibinfo {volume} {C78}},\
  \bibinfo {pages} {857} (\bibinfo {year} {2018})},\ \Eprint
  {http://arxiv.org/abs/1808.01950} {arXiv:1808.01950 [hep-ph]} \BibitemShut
  {NoStop}%
\bibitem [{\citenamefont {Jia}\ \emph {et~al.}(2019)\citenamefont {Jia} \emph
  {et~al.}}]{Jia:2019eav}%
  \BibitemOpen
  \bibfield  {author} {\bibinfo {author} {\bibfnamefont {S.}~\bibnamefont
  {Jia}} \emph {et~al.} (\bibinfo {collaboration} {Belle}),\ }\href {\doibase
  10.1103/PhysRevD.100.032006} {\bibfield  {journal} {\bibinfo  {journal}
  {Phys. Rev.}\ }\textbf {\bibinfo {volume} {D100}},\ \bibinfo {pages} {032006}
  (\bibinfo {year} {2019})},\ \Eprint {http://arxiv.org/abs/1906.00194}
  {arXiv:1906.00194 [hep-ex]} \BibitemShut {NoStop}%
\bibitem [{\citenamefont {Shen}\ \emph {et~al.}(2018)\citenamefont {Shen},
  \citenamefont {Rönchen}, \citenamefont {Meißner},\ and\ \citenamefont
  {Zou}}]{Shen:2017ayv}%
  \BibitemOpen
  \bibfield  {author} {\bibinfo {author} {\bibfnamefont {C.-W.}\ \bibnamefont
  {Shen}}, \bibinfo {author} {\bibfnamefont {D.}~\bibnamefont {Rönchen}},
  \bibinfo {author} {\bibfnamefont {U.-G.}\ \bibnamefont {Meißner}}, \ and\
  \bibinfo {author} {\bibfnamefont {B.-S.}\ \bibnamefont {Zou}},\ }\href
  {\doibase 10.1088/1674-1137/42/2/023106} {\bibfield  {journal} {\bibinfo
  {journal} {Chin. Phys.}\ }\textbf {\bibinfo {volume} {C42}},\ \bibinfo
  {pages} {023106} (\bibinfo {year} {2018})},\ \Eprint
  {http://arxiv.org/abs/1710.03885} {arXiv:1710.03885 [hep-ph]} \BibitemShut
  {NoStop}%
\bibitem [{\citenamefont {Zou}\ and\ \citenamefont
  {Hussain}(2003)}]{Zou:2002yy}%
  \BibitemOpen
  \bibfield  {author} {\bibinfo {author} {\bibfnamefont {B.~S.}\ \bibnamefont
  {Zou}}\ and\ \bibinfo {author} {\bibfnamefont {F.}~\bibnamefont {Hussain}},\
  }\href {\doibase 10.1103/PhysRevC.67.015204} {\bibfield  {journal} {\bibinfo
  {journal} {Phys. Rev.}\ }\textbf {\bibinfo {volume} {C67}},\ \bibinfo {pages}
  {015204} (\bibinfo {year} {2003})},\ \Eprint
  {http://arxiv.org/abs/hep-ph/0210164} {arXiv:hep-ph/0210164 [hep-ph]}
  \BibitemShut {NoStop}%
\bibitem [{\citenamefont {Lin}\ and\ \citenamefont {Zou}(2019)}]{Lin:2019qiv}%
  \BibitemOpen
  \bibfield  {author} {\bibinfo {author} {\bibfnamefont {Y.-H.}\ \bibnamefont
  {Lin}}\ and\ \bibinfo {author} {\bibfnamefont {B.-S.}\ \bibnamefont {Zou}},\
  }\href {\doibase 10.1103/PhysRevD.100.056005} {\bibfield  {journal} {\bibinfo
   {journal} {Phys. Rev.}\ }\textbf {\bibinfo {volume} {D100}},\ \bibinfo
  {pages} {056005} (\bibinfo {year} {2019})},\ \Eprint
  {http://arxiv.org/abs/1908.05309} {arXiv:1908.05309 [hep-ph]} \BibitemShut
  {NoStop}%
\bibitem [{\citenamefont {Faessler}\ \emph {et~al.}(2007)\citenamefont
  {Faessler}, \citenamefont {Gutsche}, \citenamefont {Lyubovitskij},\ and\
  \citenamefont {Ma}}]{Faessler:2007gv}%
  \BibitemOpen
  \bibfield  {author} {\bibinfo {author} {\bibfnamefont {A.}~\bibnamefont
  {Faessler}}, \bibinfo {author} {\bibfnamefont {T.}~\bibnamefont {Gutsche}},
  \bibinfo {author} {\bibfnamefont {V.~E.}\ \bibnamefont {Lyubovitskij}}, \
  and\ \bibinfo {author} {\bibfnamefont {Y.-L.}\ \bibnamefont {Ma}},\ }\href
  {\doibase 10.1103/PhysRevD.76.014005} {\bibfield  {journal} {\bibinfo
  {journal} {Phys. Rev.}\ }\textbf {\bibinfo {volume} {D76}},\ \bibinfo {pages}
  {014005} (\bibinfo {year} {2007})},\ \Eprint {http://arxiv.org/abs/0705.0254}
  {arXiv:0705.0254 [hep-ph]} \BibitemShut {NoStop}%
\bibitem [{\citenamefont {Dong}\ \emph {et~al.}(2009)\citenamefont {Dong},
  \citenamefont {Faessler}, \citenamefont {Gutsche}, \citenamefont
  {Kovalenko},\ and\ \citenamefont {Lyubovitskij}}]{Dong:2009yp}%
  \BibitemOpen
  \bibfield  {author} {\bibinfo {author} {\bibfnamefont {Y.}~\bibnamefont
  {Dong}}, \bibinfo {author} {\bibfnamefont {A.}~\bibnamefont {Faessler}},
  \bibinfo {author} {\bibfnamefont {T.}~\bibnamefont {Gutsche}}, \bibinfo
  {author} {\bibfnamefont {S.}~\bibnamefont {Kovalenko}}, \ and\ \bibinfo
  {author} {\bibfnamefont {V.~E.}\ \bibnamefont {Lyubovitskij}},\ }\href
  {\doibase 10.1103/PhysRevD.79.094013} {\bibfield  {journal} {\bibinfo
  {journal} {Phys. Rev.}\ }\textbf {\bibinfo {volume} {D79}},\ \bibinfo {pages}
  {094013} (\bibinfo {year} {2009})},\ \Eprint {http://arxiv.org/abs/0903.5416}
  {arXiv:0903.5416 [hep-ph]} \BibitemShut {NoStop}%
\bibitem [{\citenamefont {Dong}\ \emph {et~al.}(2010)\citenamefont {Dong},
  \citenamefont {Faessler}, \citenamefont {Gutsche},\ and\ \citenamefont
  {Lyubovitskij}}]{Dong:2009tg}%
  \BibitemOpen
  \bibfield  {author} {\bibinfo {author} {\bibfnamefont {Y.}~\bibnamefont
  {Dong}}, \bibinfo {author} {\bibfnamefont {A.}~\bibnamefont {Faessler}},
  \bibinfo {author} {\bibfnamefont {T.}~\bibnamefont {Gutsche}}, \ and\
  \bibinfo {author} {\bibfnamefont {V.~E.}\ \bibnamefont {Lyubovitskij}},\
  }\href {\doibase 10.1103/PhysRevD.81.014006} {\bibfield  {journal} {\bibinfo
  {journal} {Phys. Rev.}\ }\textbf {\bibinfo {volume} {D81}},\ \bibinfo {pages}
  {014006} (\bibinfo {year} {2010})},\ \Eprint {http://arxiv.org/abs/0910.1204}
  {arXiv:0910.1204 [hep-ph]} \BibitemShut {NoStop}%
\bibitem [{\citenamefont {Lü}\ and\ \citenamefont {Dong}(2016)}]{Lu:2016nnt}%
  \BibitemOpen
  \bibfield  {author} {\bibinfo {author} {\bibfnamefont {Q.-F.}\ \bibnamefont
  {Lü}}\ and\ \bibinfo {author} {\bibfnamefont {Y.-B.}\ \bibnamefont {Dong}},\
  }\href {\doibase 10.1103/PhysRevD.93.074020} {\bibfield  {journal} {\bibinfo
  {journal} {Phys. Rev.}\ }\textbf {\bibinfo {volume} {D93}},\ \bibinfo {pages}
  {074020} (\bibinfo {year} {2016})},\ \Eprint
  {http://arxiv.org/abs/1603.00559} {arXiv:1603.00559 [hep-ph]} \BibitemShut
  {NoStop}%
\bibitem [{\citenamefont {Xiao}\ \emph {et~al.}(2019)\citenamefont {Xiao},
  \citenamefont {Huang}, \citenamefont {Dong}, \citenamefont {Geng},\ and\
  \citenamefont {Chen}}]{Xiao:2019mst}%
  \BibitemOpen
  \bibfield  {author} {\bibinfo {author} {\bibfnamefont {C.-J.}\ \bibnamefont
  {Xiao}}, \bibinfo {author} {\bibfnamefont {Y.}~\bibnamefont {Huang}},
  \bibinfo {author} {\bibfnamefont {Y.-B.}\ \bibnamefont {Dong}}, \bibinfo
  {author} {\bibfnamefont {L.-S.}\ \bibnamefont {Geng}}, \ and\ \bibinfo
  {author} {\bibfnamefont {D.-Y.}\ \bibnamefont {Chen}},\ }\href {\doibase
  10.1103/PhysRevD.100.014022} {\bibfield  {journal} {\bibinfo  {journal}
  {Phys. Rev.}\ }\textbf {\bibinfo {volume} {D100}},\ \bibinfo {pages} {014022}
  (\bibinfo {year} {2019})},\ \Eprint {http://arxiv.org/abs/1904.00872}
  {arXiv:1904.00872 [hep-ph]} \BibitemShut {NoStop}%
\bibitem [{\citenamefont {Nieves}\ and\ \citenamefont
  {Valderrama}(2012)}]{Nieves:2012tt}%
  \BibitemOpen
  \bibfield  {author} {\bibinfo {author} {\bibfnamefont {J.}~\bibnamefont
  {Nieves}}\ and\ \bibinfo {author} {\bibfnamefont {M.~P.}\ \bibnamefont
  {Valderrama}},\ }\href {\doibase 10.1103/PhysRevD.86.056004} {\bibfield
  {journal} {\bibinfo  {journal} {Phys. Rev.}\ }\textbf {\bibinfo {volume}
  {D86}},\ \bibinfo {pages} {056004} (\bibinfo {year} {2012})},\ \Eprint
  {http://arxiv.org/abs/1204.2790} {arXiv:1204.2790 [hep-ph]} \BibitemShut
  {NoStop}%
\bibitem [{\citenamefont {Hidalgo-Duque}\ \emph {et~al.}(2013)\citenamefont
  {Hidalgo-Duque}, \citenamefont {Nieves},\ and\ \citenamefont
  {Valderrama}}]{HidalgoDuque:2012pq}%
  \BibitemOpen
  \bibfield  {author} {\bibinfo {author} {\bibfnamefont {C.}~\bibnamefont
  {Hidalgo-Duque}}, \bibinfo {author} {\bibfnamefont {J.}~\bibnamefont
  {Nieves}}, \ and\ \bibinfo {author} {\bibfnamefont {M.~P.}\ \bibnamefont
  {Valderrama}},\ }\href {\doibase 10.1103/PhysRevD.87.076006} {\bibfield
  {journal} {\bibinfo  {journal} {Phys. Rev.}\ }\textbf {\bibinfo {volume}
  {D87}},\ \bibinfo {pages} {076006} (\bibinfo {year} {2013})},\ \Eprint
  {http://arxiv.org/abs/1210.5431} {arXiv:1210.5431 [hep-ph]} \BibitemShut
  {NoStop}%
\end{thebibliography}%

\end{document}